\documentclass[journal]{IEEEtran}
\usepackage{graphicx}
\usepackage{cite}
\usepackage{amsmath,amssymb,amsfonts}
\usepackage{algorithmic}
\usepackage{textcomp}
\usepackage{booktabs}
\usepackage{xspace}
\usepackage{svg}
\usepackage{amsmath}
\usepackage{footnote}
\usepackage{makecell}
\usepackage{multirow}
\usepackage{enumitem}
\usepackage{array}
\usepackage[nolist,nohyperlinks]{acronym}
\usepackage{hyperref}
\usepackage{subcaption}
\makesavenoteenv{tabular}
\usepackage{caption, subcaption}
\usepackage[acronym]{glossaries}
\acrodef{CRC}{Colorectal cancer}
\acrodef{GI}{Gastrointestinal}
\acrodef{CADx}{Computer-Aided diagnosis}
\acrodef{CNN}{Convolutional Neural Network}
\acrodef{FPS}{Frame Per Second}
\acrodef{DSC}{Dice Coefficient}
\acrodef{SOTA}{state-of-the-art}
\acrodef{mIoU}{mean Intersection over Union}
\begin{document}
\title{Automatic Polyp Segmentation with Multiple Kernel Dilated Convolution Network}
\author{
\IEEEauthorblockN{Nikhil Kumar Tomar\IEEEauthorrefmark{1}, Abhishek Srivastava\IEEEauthorrefmark{2},
Ulas Bagci\IEEEauthorrefmark{3},
Debesh Jha\IEEEauthorrefmark{3}
}\\
\IEEEauthorblockA{\IEEEauthorrefmark{1}School of Computer and Information Sciences, Indira Gandhi National Open University\\ 
\IEEEauthorrefmark{2}Computer Vision and Pattern Recognition Unit, Indian Statistical Institute\\
\IEEEauthorrefmark{3} Machine and Hybrid Intelligence Lab, Department of Radiology, Northwestern University, USA\\ 
}
}
\maketitle
\begin{abstract}
The detection and removal of precancerous polyps through colonoscopy is the primary technique for the prevention of colorectal cancer worldwide. However, the miss rate of colorectal polyp varies significantly among the endoscopists. It is well known that a computer-aided diagnosis (CAD) system can assist endoscopists in detecting colon polyps and minimize the variation among endoscopists. In this study, we introduce a novel deep learning architecture, named {\textbf{MKDCNet}}, for automatic polyp segmentation robust to significant changes in polyp data distribution. MKDCNet is simply an encoder-decoder neural network that uses the pre-trained \textit{ResNet50} as the encoder and novel \textit{multiple kernel dilated convolution (MKDC)} block that expands the field of view to learn more robust and heterogeneous representation. Extensive experiments on four publicly available polyp datasets and cell nuclei dataset show that the proposed MKDCNet outperforms the state-of-the-art methods when trained and tested on the same dataset as well when tested on unseen polyp datasets from different distributions. With rich results, we demonstrated the robustness of the proposed architecture. From an efficiency perspective, our algorithm can process at ($\approx45$) frames per second on RTX 3090 GPU. MKDCNet can be a strong benchmark for building real-time systems for clinical colonoscopies. The code of the proposed MKDCNet is available at \url{https://github.com/nikhilroxtomar/MKDCNet}. 
\end{abstract}


\section{Introduction}
\acf{CRC} is the second leading cause of cancer-related death and the third leading common cause of cancer worldwide~\cite{sung2021global}.  The five-year survival rate is 90\% for 39\% of the patients that are diagnosed with localized stage disease but declines to 71\% and 14\% once diagnosed with regional and distant stage respectively~\cite{american2020colorectal}. Colonoscopy is considered the primary technique for colon cancer screening because it offers both detecting and removal of the polyp in a single operation. U.S. Preventive Services Task Force recommends forty-five to be considered as the new fifty for screening of \ac{CRC}~\cite{davidson2021screening}. Colonoscopy can reduce the mortality through early detection at treatable stage and remove precancerous adenomas~\cite{zauber2012colonoscopic,brenner2011protection}. 

During the colonoscopy operation, the average miss rate of the polyp is around 22-28\%~\cite{leufkens2012factors}. It is mainly because colonoscopy is an operator-dependent procedure and high inter-observer variations are seen in endoscopists' skills in detecting polyps~\cite{hetzel2010variation}. During routine colonoscopy, the most frequently missed polyps are flat and smaller polyps~\cite{heresbach2008miss,short2015colorectal,wang2018development}. Studies have shown that even a 1\% increase in adenomas detection leads to a 3\% decrease in the risk of interval colon cancer~\cite{corley2014adenoma}. Therefore, it is highly critical to decrease the polyp miss-rate via an automated systems for \ac{CRC} screening.


A \ac{CADx} can highlight the suspicious frames and improve colonoscopy procedures. Jha et al.~\cite{jha2020doubleu} proposed DoubleU-Net that used two U-Net's where the output of first U-Net acts as a soft-attention to the other. The network uses VGG-19 as an encoder and efficient blocks such as squeeze and excitation network~\cite{hu2018squeeze} and atrous spatial pyradimal pooling~\cite{chen2017deeplab} to capture some semantically meaningful information. DoubleU-Net showed \ac{SOTA} results on different biomedical image segmentation datasets. Wu et al.~\cite{wu2022polypseg+} proposed a lightweight context-aware network, PolypSeg+, for real-time polyp segmentation. The proposed architecture can capture distinguishable polyp features even with less trainable parameters and retain real-time speed. Tomar et al.~\cite{tomar2022fanet} proposed a feedback attention network (FANet) for improved biomedical image segmentation, where they showed the \ac{SOTA} performance on seven publicly available benchmark datasets. FANet unifies the mask of the previous epoch with the current training epoch and rectifies the prediction iteratively during the test time for improved performance. Ji et al.~\cite{ji2021progressively} proposed a progressively normalized self-attention network (PNS-Net) for video polyp segmentation. Shen et al.~\cite{shen2021hrenet} proposed a hard region enhancement network (HRENet) for automatic polyp segmentation.

\begin{figure*}[t!]
    \centering
    \includegraphics[width=0.85\textwidth]{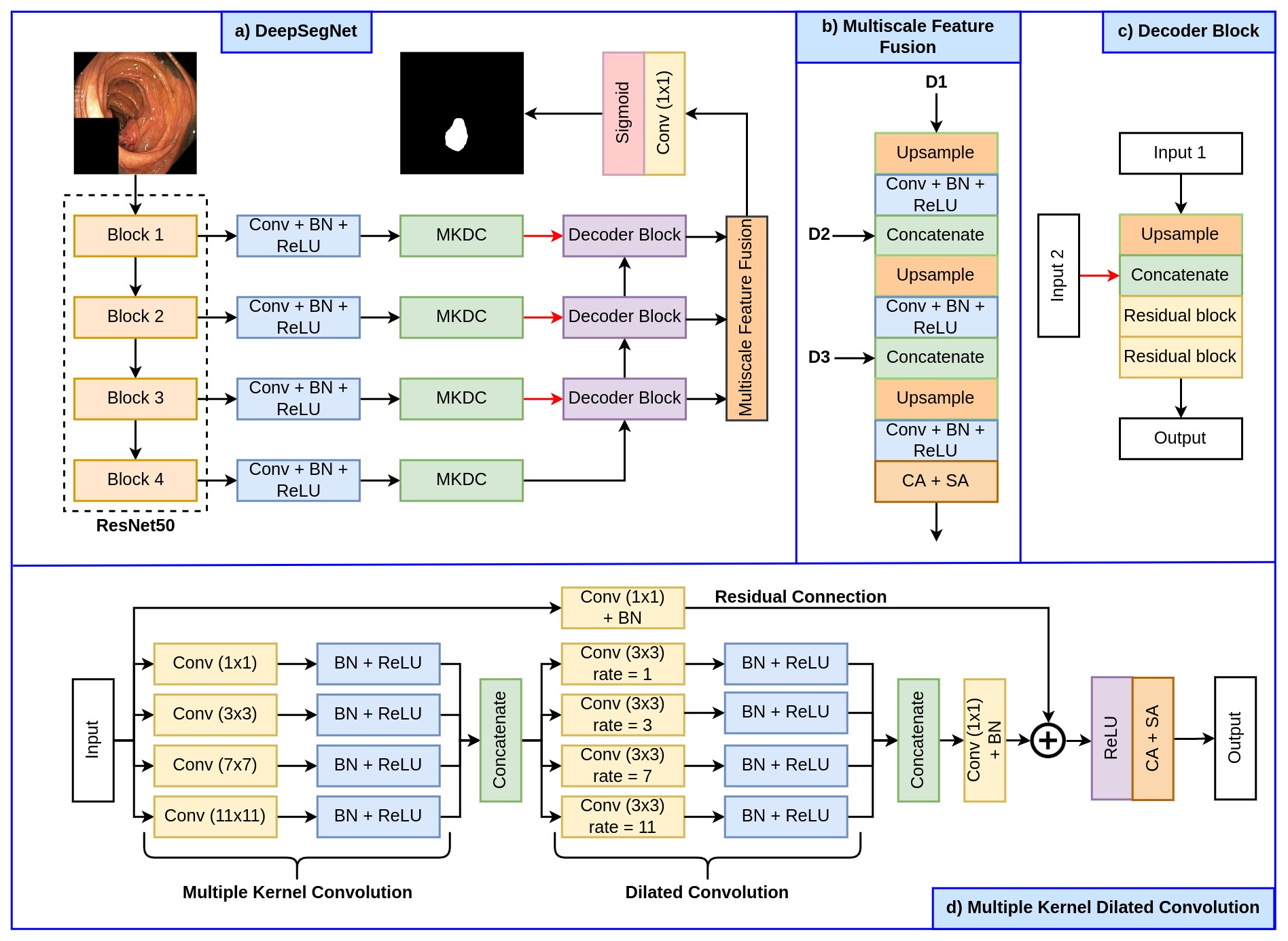}
    \caption{Block diagram of the proposed MKDCNet along with its building blocks.}
    \label{fig:proposed-architecture}
\end{figure*}

Despite the several automated methods proposed to improve the accuracy of polyp segmentation, further investigations are required to show the generalizability of the existing and the proposed method. Currently, most of the algorithms are only trained and tested on the same datasets~\cite{ji2021progressively,jha2019resunet++,zhou2018unet++,tomar2022fanet,fan2020pranet}. Therefore, we aim to develop a novel deep learning algorithm to work well on varying distribution datasets coming from different institutions across different countries. To this end, we introduce a multiple kernel dilated convolution network (MKDCNet) architecture and test its performance on four still image datasets (polyps) and one cell nuclei dataset.

The main contribution of our work can be summarized as follows:
\begin{enumerate}
    \item We present a novel deep learning architecture, MKDCNet, that utilizes novel multiple kernel dilated convolution block to increase the field of view of convolution kernel in order to capture local and global features. The multiscale feature fusion block fuses different decoder blocks output for more robust feature representation that helps in accurate polyp segmentation. 
    
    \item We obtained \ac{SOTA} results on four publicly available polyp datasets (same train-test set), and a nuclei segmentation dataset. Similarly, the proposed method outperformed other methods on three cross-center polyp dataset. Extensive experimental results shows the strong learning and generalization ability of MKDCNet. 
\end{enumerate}

\vspace{-5mm}

\section{Method}
The proposed MKDCNet architecture is illustrated in Figure~\ref{fig:proposed-architecture}. The architecture begins with a pre-trained ResNet50~\cite{he2016deep} as the encoder from which we extract four different feature maps. Each of these feature map is then passed through a sequence of $3\times3$ convolution layer, batch normalization, and a ReLU activation function. The output from the ReLU activation function is then passed through our novel Multiple Kernel Dilated Convolution (MKDC) block, which consists of multiple parallel convolution layers with different kernel sizes and dilation rates. After that, we have three decoder blocks, the output from all the three decoder blocks is passed through a Multiscale Feature Fusion (MSFF) block where we upsample and fuse the feature map to produce a more robust semantic representation. Finally, this feature map is then passed through a $1\times1$ convolution followed by a sigmoid activation function generating a binary segmentation mask.

\subsection{Multiple kernel dilated convolution (MKDC) block}
The \textit{MKDC} block begins with four parallel convolution layers with a kernel size of $1\times1$, $3\times3$, $7\times7$ and $11\times11$ respectively. The kernel size's progressive increase helps capture a broad range of features, allowing the network to learn a more robust representation. Each convolution layer is then followed by batch normalization and a ReLU activation function. Next, each of these feature maps are then concatenated and passed through four parallel convolution layer, each having a dilation rate of $1$, $3$, $7$ and $11$, respectively. The use of different dilated convolutions helps to further expand the field of view and allows the network to capture more details and refine the significant features. In this sense, the \textit{MKDC} is similar to multi-resolution strategies but in our case we capture rich details with convolutional kernels instead of using multiple parallel architectures or iterative and simultaneous connection from each resolutions. Each of the convolution layer is then followed by batch normalization and ReLU activation function.  After that, we perform a concatenation over these features and feed them to a $1\times1$ convolution followed by a residual connection. Finally, the generated feature maps are passed through a channel and spatial attention mechanism which further highlight the significant features.

\subsection{Decoder block}
The decoder block begins with a bilinear upsampling which increases the spatial dimensions (height and width) of the input feature map by a factor of two. After that, the upsampled feature map is then concatenated with the output of another \textit{MKDC} block, that brings more semantic information to the decoder increasing its feature representation. Next, we have two residual block, where each residual block consists of a convolutional block and an identity mapping connecting the input and output of the convolutional block. The convolutional block begins with two $3\times3$ convolution layer, where each is followed by a batch normalization and a ReLU activation function.

\subsection{Multiscale feature fusion (MSFF) block}
We use the proposed \textit{MSFF} block to enhance the feature at different scales by aggregating them to produce a more robust feature representation. The \textit{MSFF} block takes the output from the first decoder block and passes it through a bilinear upsampling layer to increase its spatial dimensions by a factor of two. After that, it is followed by a $3\times3$ convolution layer, batch normalization and a ReLU activation function. The output of the ReLU activation function is then concatenated with the output from the second decoder block. Next, we again follow a bilinear upsampling layer where the concatenated feature map is upsampled by a factor of two and then followed by a $3\times3$ convolution layer, batch normalization and a ReLU activation function. The output from the ReLU activation function is then concatenated with the output from the third decoder block. After this, the feature map is again upsampled and passed through a $3\times3$ convolution layer, batch normalization and a ReLU activation function. The feature map is then passed through channel and spatial attention mechanism that focus on significant features and thus improve the feature representation and its robustness.

\begin{table} [t!]
\caption{Details of the datasets used in our experiments.}
\centering
\begin{tabular}{@{}l|l|l|l@{}} 
\toprule
\textbf{Dataset} &\textbf{Images} &\textbf{Size} &\textbf{Application}\\ 
\midrule
Kvasir-SEG~\cite{jha2020kvasir} & $1000$ & Variable &Colonoscopy\\
BKAI-IGH~\cite{lan2021neounet} & $1000$ & $1280 \times 995$ &Colonoscopy\\ 
CVC-ClinicDB~\cite{bernal2015wm} & $612$ & $384 \times 288$  &Colonoscopy\\ 
MedAI Challenge test set~\cite{hicks2021medai} &200 & Variable &Colonoscopy\\
2018 Data Science Bowl~\cite{caicedo2019nucleus} & 670 & $256\times 256$ & Nuclie \\ 

\bottomrule
\end{tabular}
\label{table:datasettable}
\end{table}	

\begin{table*}[t!]
\centering
\caption{Quantitative results on the experimented datasets.}
 \begin{tabular} {@{}l|c|c|c|c|c|c|c@{}}
\toprule
\textbf{Method} &\textbf{DSC} & \textbf{mIoU} &\textbf{Rec.}  &\textbf{Prec.}& \textbf{Acc.} &\textbf{F2} &\textbf{FPS}\\ 
\hline
\multicolumn{8}{@{}l}{\textbf{Dataset: Kvasir-SEG~\cite{jha2020kvasir}}}                   \\ \hline
U-Net\cite{ronneberger2015u}	&0.8264	&0.7472	&0.8504	&0.8703	&0.9510	&0.8353	&156.83\\
ResU-Net\cite{zhang2018road}&	0.7642	&0.6634	&0.8025	&0.8200	&0.9341	&0.7740	&\textbf{196.85}\\
U-Net++~\cite{zhou2018unet++}&0.8228	&0.7419	&0.8437	&0.8607	&0.9491	&0.8295	&126.14\\
ResU-Net++~\cite{jha2019resunet++}&	0.6453	&0.5341	&0.6964	&0.7080	&0.9044	&0.6575	&57.99\\
HarDNet-MSEG~\cite{huang2021hardnet}	&0.8260	&0.7459	&0.8485	&0.8652	&0.9492	&0.8358	&42.00\\
DeepLabV3+ (ResNet50)~\cite{chen2018encoder} &0.8837	&0.8173	&0.9014	&0.9028	&\textbf{0.9679}	&0.8904	&102.62\\
DDANet~\cite{tomar2021ddanet}	&0.7415	&0.6448	&0.7953	&0.7670	&0.9326	&0.7640	&88.70\\
\textbf{MKDCNet (Ours)}	&\textbf{0.8887}	&\textbf{0.8267}	&\textbf{0.9076}	&\textbf{0.9088}	&0.9677	&\textbf{0.8954}	&47.54\\

\midrule
\multicolumn{8}{@{}l}{\textbf{Dataset: BKAI-IGH~\cite{lan2021neounet}}}                   \\ \hline

U-Net~\cite{ronneberger2015u}	&0.8286	&0.7599	&0.8295	&0.8999	&0.9903	&0.8264	&\textbf{160.27}\\
ResU-Net\cite{zhang2018road}	&0.7433	&0.6580	&0.7447	&0.8711	&0.9843	&0.7387	&128.93\\
U-Net++~\cite{zhou2018unet++}	&0.8275	&0.7563	&0.8388	&0.8942	&0.9895	&0.8308	&123.45\\
ResU-Net++~\cite{jha2019resunet++}	&0.7130	&0.6280	&0.7240	&0.8578	&0.9832	&0.7132	&55.86\\
HarDNet-MSEG~\cite{huang2021hardnet}	&0.7627	&0.6734	&0.7532	&0.8344	&0.9863	&0.7528	&41.20\\
DeepLabV3+ (ResNet50)~\cite{chen2018encoder}	&0.8937	&0.8314	&0.8870	&0.9333	&\textbf{0.9937}	&0.8882	&99.16\\
DDANet~\cite{tomar2021ddanet}	&0.7269	&0.6507	&0.7454	&0.7575	&0.9851	&0.7335	&86.46\\
\textbf{MKDCNet (Ours)}	& \textbf{0.8978}	&\textbf{0.8392}	&\textbf{0.8955}	&\textbf{0.9365}	&0.9934	&\textbf{0.8947}	&45.98\\

\midrule

\multicolumn{8}{@{}l}{\textbf{Dataset: 2018 Data Science Bowl~\cite{caicedo2019nucleus}}}                   \\ \hline

U-Net~\cite{ronneberger2015u}	&0.9122	&0.8476	&0.9021	&\textbf{0.9339}	&0.9799	&0.9052	&160.53\\
ResU-Net~\cite{zhang2018road}	&0.9183	&0.8546	&0.9236	&0.9198	&0.9809	&0.9207	&\textbf{188.74}\\
U-Net++~\cite{zhou2018unet++}	&0.9114	&0.8479	&0.9107	&0.9269	&0.9799	&0.9101	&119.45\\
ResU-Net++~\cite{jha2019resunet++}	&0.9157	&0.8508	&0.9162	&0.9211	&0.9798	&0.9153	&55.91\\
HarDNet-MSEG~\cite{huang2021hardnet} &0.8344 &0.7327 &0.8686 &0.8251 &0.9640 &0.8538 &40.53 \\
DeepLabV3+ (ResNet50)~\cite{chen2018encoder}	&0.9027	&0.8306	&0.9220	&0.8902	&0.9774	&0.9134	&98.53\\
DDANet~\cite{tomar2021ddanet}	&0.9117	&0.8452	&0.8452	&0.9297	&0.9792	&0.9053	&90.33\\
\textbf{MKDCNet (Ours)}	&\textbf{0.9204}	&\textbf{0.8586}	&\textbf{0.9270}	&0.9194	&\textbf{0.9815}	 &\textbf{0.9237}	&46.56\\
\bottomrule
\end{tabular}
\label{tab:results}
\end{table*}

\section{Experimental setup}
In this section, we will present the datasets, evaluation metrics, and implementation details used in this study.

\begin{table*}[t!]
\centering
\caption{Quantitative results on the unseen polyp dataset.}
 \begin{tabular} {@{}l|c|c|c|c|c|c|c@{}}
\toprule
\textbf{Method} &\textbf{DSC} & \textbf{mIoU} &\textbf{Rec.}  &\textbf{Prec.}& \textbf{Acc.} &\textbf{F2} &\textbf{FPS}\\ 
\hline

\multicolumn{8}{@{}l}{\textbf{Train Dataset: Kvasir-SEG\cite{jha2020kvasir}, Test Data: Unseen CVC-ClinicDB~\cite{bernal2015wm}}}                   \\ \hline
U-Net~\cite{ronneberger2015u}	&0.6336	&0.5433	&0.6982	&0.7891	&0.9484	&0.6563	&166.05\\
ResU-Net~\cite{zhang2018road}	&0.5970	&0.4967	&0.6210	&0.8005	&0.9465	&0.5991	&\textbf{195.38}\\
U-Net++~\cite{zhou2018unet++}	&0.6350	&0.5475	&0.6933	&0.7967	&0.9504	&0.6556	&127.80\\
ResU-Net++~\cite{jha2019resunet++}	&0.4642	&0.3585	&0.5880	&0.5770	&0.9159	&0.5084	&57.96\\
HarDNet-MSEG~\cite{huang2021hardnet}	&0.6960	&0.6058 &0.7173	&0.8528	&0.9592	&0.7010	&42.38\\
DeepLabV3+ (ResNet50)~\cite{chen2018encoder}	&0.8142	&0.7388	&0.8331	&\textbf{0.8735}	&\textbf{0.9717}	&0.8198	&103.17\\
DDANet\cite{tomar2021ddanet}	&0.5234	&0.4183	&0.6502	&0.5935	&0.9275	&0.5718	&91.32\\
\textbf{MKDCNet (Ours)}	&\textbf{0.8243}	&\textbf{0.7466}	&\textbf{0.8494}	&0.8637	&0.9709	&\textbf{0.8325}	&46.71\\
\midrule

\multicolumn{8}{@{}l}{\textbf{Train Dataset: Kvasir-SEG\cite{jha2020kvasir}, Test Data: Unseen BKAI-IGH~\cite{lan2021neounet}}}  
 \\ \hline
U-Net~\cite{ronneberger2015u} &0.6347	&0.5686	&0.6986	&0.7882	&0.9753	&0.6591	&162.60 \\
ResU-Net~\cite{zhang2018road} &0.5836	&0.4931	&0.6716	&0.6549	&0.9671	&0.6177	&\textbf{199.02} \\
U-Net++~\cite{zhou2018unet++} &0.6269	&0.5592	&0.6900	&0.7968	&0.9741	&0.6493	&128.59 \\
ResU-Net++~\cite{jha2019resunet++} &0.4166	&0.3204	&0.6979	&0.3922	&0.9061	&0.5019	&57.22 \\
HarDNet-MSEG~\cite{huang2021hardnet} &0.6502	&0.5711	&0.7420	&0.7469	&0.9713	&0.6830	&42.44 \\
DeepLabV3+ (ResNet50)~\cite{chen2018encoder} &0.7286	&0.6589	&0.7919	&0.8123	&\textbf{0.9787}	&0.7493	&103.25 \\
DDANet\cite{tomar2021ddanet} &0.5006	&0.4115	&0.6612	&0.4825	&0.9507	&0.5592	&91.73 \\
\textbf{MKDCNet (Ours)} &\textbf{0.7483}	&\textbf{0.6782}	&\textbf{0.8087}	&\textbf{0.8155}	&0.9756	&\textbf{0.7651}	&42.741 \\

\midrule
\multicolumn{8}{@{}l}{\textbf{Train Dataset: Kvasir-SEG\cite{jha2020kvasir}, Test Data: MedAI Challenge test data (polyp)~\cite{hicks2021medai}}}                   \\ \hline
U-Net~\cite{ronneberger2015u} & 0.6716	& 0.5725 & 0.7462 & 0.7438 & 0.9279	 &0.6957	 &159.90 \\
ResU-Net~\cite{zhang2018road} &0.6165	&0.4991	&0.6726	&0.6977	&0.9139	&0.6315	&\textbf{192.90}\\
U-Net++~\cite{zhou2018unet++}  &0.6638	&0.5702	&0.7258	&0.7594	&0.9333	&0.6845	&128.64 \\
ResU-Net++~\cite{jha2019resunet++}  &0.4306	&0.3246	&0.5865	&0.4677	&0.8629	&0.4793	&60.20 \\
HarDNet-MSEG~\cite{huang2021hardnet}   &0.6821	&0.5877	&0.756	&0.7689	&0.9271	&0.7006	&43.91    \\
DeepLabV3+ (ResNet50)~\cite{chen2018encoder} &0.7784	&0.6875	&0.8332	&0.8054	&\textbf{0.9544}	&0.7989	&106.77\\
DDANet~\cite{tomar2021ddanet} &0.5738	&0.4643	&0.6638	&0.6131	&0.9141	&0.6058	&90.22 \\            
\textbf{MKDCNet (Ours)} &\textbf{0.7961}	&\textbf{0.7054}	&\textbf{0.8397}	&\textbf{0.8151}&0.9532	&\textbf{0.8103}	&46.59  \\

\midrule
\multicolumn{8}{@{}l}{\textbf{Train Dataset: BKAI-IGH~\cite{lan2021neounet}, Test Data: MedAI Challenge test data (polyp)~\cite{hicks2021medai}}}                   \\ \hline
U-Net~\cite{ronneberger2015u} &0.5840	&0.4837	&0.5925	&0.8147	&0.9155	&0.5726	&166.94 \\
ResU-Net~\cite{zhang2018road} & 0.4620	&0.3605	&0.4822	&0.6989	&0.8930	&0.4525	&\textbf{196.09}\\
U-Net++~\cite{zhou2018unet++}  &0.5554	&0.4530	&0.6037	&0.7475	&0.8941	&0.5591	&126.01\\
ResU-Net++~\cite{jha2019resunet++}  &0.3288	&0.2419	&0.3560	&0.4779	&0.8666	&0.3313	&59.25\\
HarDNet-MSEG~\cite{huang2021hardnet} & 0.4466	&0.3550	&0.4204	&0.7427	&0.9017	&0.4210	&42.84  \\
DeepLabV3+ (ResNet50)~\cite{chen2018encoder} &0.6541	&0.5675	&0.6711	&0.8535	&0.9284	&0.6552	&100.86\\
DDANet~\cite{tomar2021ddanet} &0.5322	&0.4281	&0.5764	&0.6547	&0.8952	&0.5351	&91.02 \\            
\textbf{MKDCNet (Ours)} &\textbf{0.6985}	&\textbf{0.6078}	&\textbf{0.7210}	&\textbf{0.8360}	&\textbf{0.9366}	&\textbf{0.7009}&48.05 \\

\bottomrule
\end{tabular}
\label{tab:resultsunseen}
\end{table*}

\begin{figure*}[t!]
    \centering
    \includegraphics[width=0.8\textwidth]{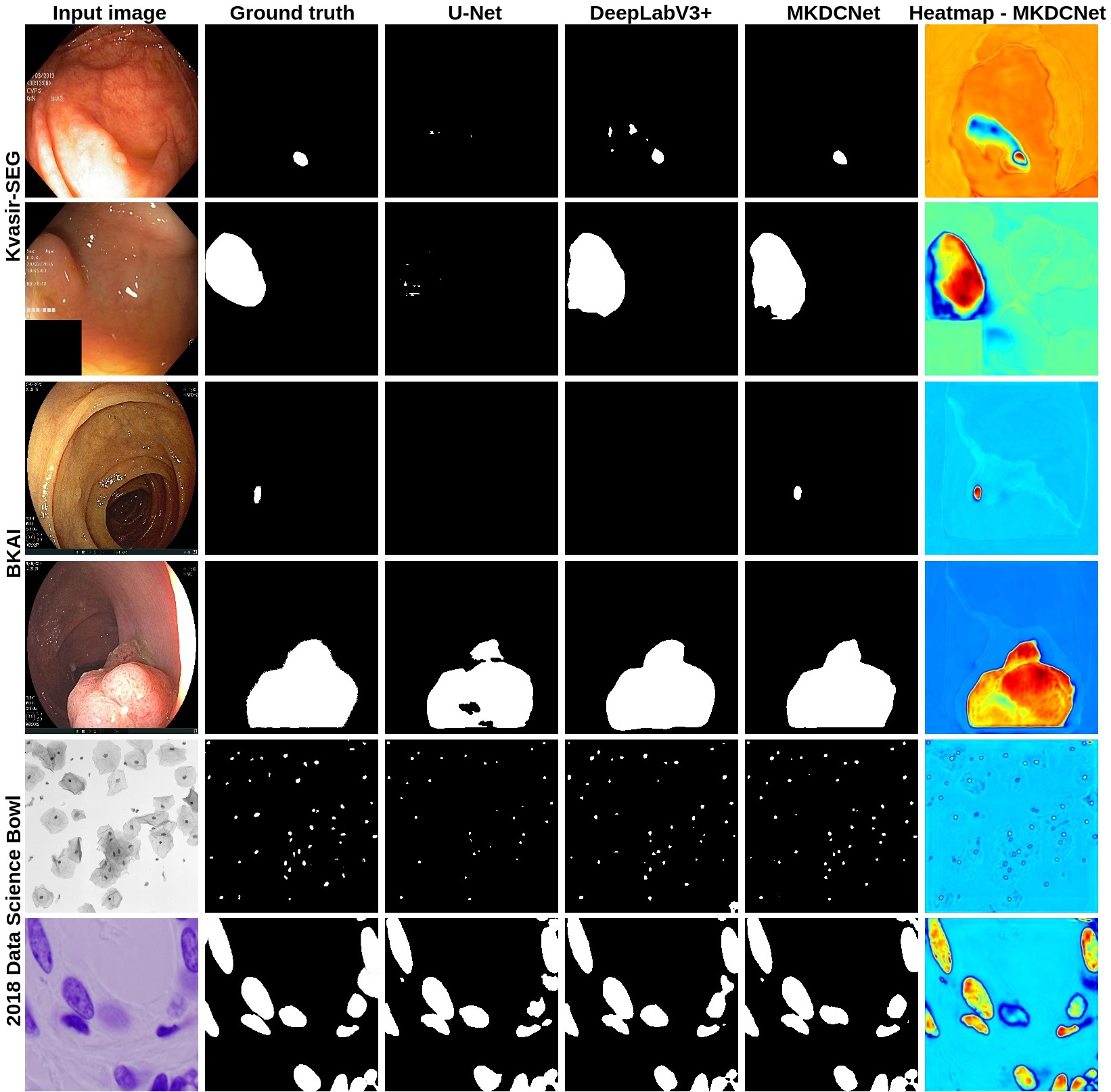}
    \caption{Qualitative results comparison along with the heatmap on the Kvasir-SEG~\cite{jha2020kvasir}, BKAI-IGH~\cite{lan2021neounet}, and 2018 Data Science Bowl~\cite{caicedo2019nucleus} datasets.  The heatmaps provide insight into the intermediate feature maps from the multi scale feature fusion block.  The heatmap shows the \textit{region of interest} and its statistical significance and the color intensity shows the \textit{effect}. The \textit{red} and \textit{yellow} colors denote the most significant feature and the \textit{blue} color denote the least significance feature.}
    \label{fig:qualitative-results}
\end{figure*}

\subsection{Datasets and evaluation}
For this study, we have select four publicly available polyp datasets and a nuclie segmentation dataset. The details about the number of images, their size, and their application can be found in Table~\ref{table:datasettable}. We have utilized Kvasir-SEG~\cite{jha2020kvasir}, BKAI-IGH~\cite{lan2021neounet}, CVC-ClinicDB~\cite{bernal2015wm}, and MedAI challenge test set~\cite{hicks2021medai} datasets for the polyp segmentation task. For the cell nuclei segmentation task, we have used the 2018 Data Science Bowl~\cite{caicedo2019nucleus} dataset.  To evaluate the performance of all the models, we have used metrics such as \ac{DSC},  \ac{mIoU}, precision, recall, accuracy, F2-score, and \ac{FPS}. 

\subsection{Implementation details}
We have implemented the proposed MKDCNet and the SOTA methods using the PyTorch framework. For a fair comparison, we have used the same set of hyperparameters for all models used in this study. All models were trained on NVIDIA RTX 3090 GPU, where both the images and masks were first resized to $256 \times 256$ pixels for better utilization of GPU. The datasets were then split into training, validation and testing in the ratio of 80:10:10, except for Kvasir-SEG, where a split of 880/120 was used for training and testing respectively. An online data augmentation strategy was used on the training dataset which includes random rotation, horizontal flipping, vertical flipping and coarse dropout. The data augmentation helped to increase the robustness of the model. All the models were trained with an Adam optimizer having a learning rate of 1e$^{-4}$  with a batch size of 16. A combination of binary cross-entropy loss and dice loss was used. ReduceLROnPlateau was used while training to reduce the learning rate for better performance. An early stopping criterion was also used to stop the training when the model stops improving.

\begin{table*} [t!]
\caption{Ablation study of the proposed MKDCNet on the Kvasir-SEG~\cite{jha2020kvasir}.}
\centering
\begin{tabular}{@{}l|l|l|l|l|l@{}} 
\toprule
\textbf{No.} &\textbf{Method} &\textbf{DSC} &\textbf{mIoU} & \textbf{Recall} & \textbf{Precision}\\ 
\midrule
\#1 & MKDCNet w/o Multiple Kernel Dilated Convolution & 0.8763	&0.8138	&0.8997	&0.9071 \\
\#2 & MKDCNet w/o Multiscale Feature Fusion &0.8720	&0.8045	&0.8974	&0.8931 \\
\#3 & MKDCNet w/o Multiple Kernel Dilated Convolution \& Multiscale Feature Fusion &0.8785	&0.8073	&0.9003	&0.8953 \\
\#4 & MKDCNet &\textbf{0.8887}	&\textbf{0.8267}	&\textbf{0.9076}	&\textbf{0.9088} \\
\bottomrule
\end{tabular}
\label{table:ablation-study-table}
\vspace{-6mm}
\end{table*}	

\section{Result}
At first, we performed validation of the algorithms on same datasets (same distribution). Next, we tested the trained model on completely unseen polyp datasets from different medical centers (different distribution). 

\subsection{Performance test on the same dataset}
Table~\ref{tab:results} shows the result of the MKDCNet and \ac{SOTA} methods. On the Kvasir-SEG dataset, MKDCNet achieved a \ac{DSC} of 0.8887 and \ac{mIoU} of 0.8267 and outperformed the most competitive benchmarking method DeepLabv3+ with ResNet50 encoder with a margin of 0.5\%  in DSC and 0.94\% in mIoU. Similarly, MKDCNet had a higher recall, precision, F2-score and nearly equal accuracy. Both DeepLabv3+ and MKDCNet had a real-time speed. Similarly, with the BKAI-IGH~\cite{lan2021neounet}, our method outperformed DeepLabv3+ with a margin of 0.41\% in \ac{DSC} and 0.78\% in \ac{mIoU}. Additionally, we performed experiments on the 2018 Data Science Bowl~\cite{caicedo2019nucleus} dataset, where we showed that our method consistently outperforms all other baseline methods. Figure~\ref{fig:qualitative-results} showed the example of qualitative results along with the heatmap. The qualitative results indicated that MKDCNet had better segmentation as compared to the UNet~\cite{ronneberger2015u} and DeepLabv3+~\cite{chen2018encoder}. 

\subsection{Performance test on completely unseen dataset}
Table~\ref{tab:resultsunseen} shows the results on the unseen dataset. For the unseen CVC-ClinicDB~\cite{bernal2015wm}, our MKDCNet outperformed DeepLabv3+ with 1.01\% in \ac{DSC} and 0.78\% in \ac{mIoU} showing the superior generalization capability of our proposed method compared to others. Similarly, for the unseen BKAI-IGH dataset~\cite{lan2021neounet}, our method outperformed best performing DeepLabv3+ by 1.97\% in \ac{DSC} and 1.93\% in \ac{mIoU}. For MedAI challenge test dataset, we only evaluated the performance on 200 positive polyp images provided by the task organizers. The models trained on Kvasir-SEG obtained better performance on the MedAI challenge dataset and slightly weaker performance with the BKAI datasets, which might be because BKAI-IGH dataset was captured at a different hospital (Institute of Gastroenterology and Hepatology (IGH), Vietnam), whereas the MedAI challenge dataset came from the HyperKvasir~\cite{borgli2020hyperkvasir} whose distribution was similar to Kvasir-SEG (as both of them are captured at Vestre Viken Hospital Trust, Norway), despite the image frames being different. For both models trained on Kvasir-SEG and BKAI-IGH, proposed MKDCNet outperformed DeepLabv3+ by 1.77\% and 4.4\% in \ac{DSC}, respectively. 

\vspace{-6mm}

\subsection{Ablation study}
In Table~\ref{table:ablation-study-table}, we presented the ablation study on Kvasir-SEG dataset. When we compared setting \#3 and setting \#4, there was a 1.02\% improvement in \ac{DSC} and a 1.94\% in \ac{mIoU} with the multiple Kernel dilated convolution and multiscale feature fusion block in the network.  Similarly, the Table~\ref{table:ablation-study-table} also showed an improvement over both of the individual blocks.

\section{Conclusion}
We presented a novel architecture, MKDCNet, that utilizes  ResNet50 as an encoder and the novel multiple kernel dilated convolution block to learn more robust representation to automatically segment polyps from colonoscopy images with high performance. Extensive experimental results on four publicly available datasets (both on the same set as well as on completely unseen datasets) consistently showed that MKDCNet has the promising capability to improve the segmentation accuracy. With MKDCNet, we obtained a real-time processing speed of nearly 45 frames per second. Our results exhibited that MKDCNet has a better generalizability, accuracy, and real-time speed. Thus, MKDCNet can be a strong new baseline for developing artificial intelligence-based support to improve the traditional colonoscopy procedure. In the future work, we plan to exploit MKDCNet under federated learning settings where we can train multiple institute datasets and minimize the privacy concerns raised by each center. 
\subsubsection*{Acknowledgement}
This project is supported by the NIH funding: R01-CA246704 and R01-CA240639.

\bibliographystyle{IEEEtran}
\bibliography{references} 

\ifCLASSOPTIONcaptionsoff
  \newpage
\fi
\vfill
\end{document}